# Chaos and COSMOS—Considerations on QSM methods with multiple and single orientations and effects from local anisotropy

Dimitrios G. Gkotsoulias[1], Carsten Jäger[2], Roland Müller[1], Tobias Gräßle[3], Karin M. Olofsson[4], Torsten Møller[5], Steve Unwin[6], Catherine Crockford[7,8,9], Roman M. Wittig[7,8,9], Berkin Bilgic[10,11,12], Harald E. Möller[1]

[1] *Nuclear Magnetic Resonance Methods & Development Group, Max Planck Institute for Human Cognitive and Brain Sciences, Leipzig, Germany*

[2] *Department of Neurophysics, Max Planck Institute for Human Cognitive and Brain Sciences, Leipzig, Germany*

[3] *Epidemiology of Highly Pathogenic Microorganisms, Robert Koch-Institute, Berlin, Germany*

[4] *National Veterinary Institute, Uppsala, Sweden*

[5] *Kolmården Wildlife Park, Norrköping, Sweden*

[6] *Wildlife Health Australia, Canberra, Australia*

[7] *Department of Human Behavior, Ecology and Culture, Max Planck Institute for Evolutionary Anthropology, Leipzig, Germany*

[8] *The Ape Social Mind Lab, Institut des Sciences Cognitives Marc Jeannerod, Bron, France*

[9] *Taï Chimpanzee Project, Centre Suisse de Recherches Scientifiques, Abidjan, Côte d'Ivoire*

[10] *Athinoula A. Martinos Center for Biomedical Imaging, Massachusetts General Hospital, Boston, MA, USA*

[11] *Harvard Medical School, Boston, MA, USA*

[12] *Harvard/MIT Health Sciences and Technology, Cambridge, MA, United States*

**Corresponding author:**

Dimitrios G. Gkotsoulias
Max Planck Institute for Human Cognitive and Brain Sciences
Stephanstraße 1a, 04103 Leipzig, Germany
gkotsoulias@cbs.mpg.de

**Word count:**

| | |
|---|---|
| Text: | 2,593 words (without legends, references and statements) |
| Abstract: | 239 words |
| References: | 33 |
| Figures: | 4 |
| Tables: | 1 |

**Manuscript type:**

Technical Note

# Abstract

**Purpose:** Field-to-susceptibility inversion in quantitative susceptibility mapping (QSM) is ill-posed and needs numerical stabilization through either regularization or oversampling by acquiring data at three or more object orientations. Calculation Of Susceptibility through Multiple Orientations Sampling (COSMOS) is an established oversampling approach and regarded as QSM gold standard. It achieves a well-conditioned inverse problem, requiring rotations by 0°, 60° and 120° in the *yz*-plane. However, this is impractical *in vivo*, where head rotations are typically restricted to a range of ±25°. Non-ideal sampling degrades the conditioning with residual streaking artifacts whose mitigation needs further regularization. Moreover, susceptibility anisotropy in white matter is not considered in the COSMOS model, which may introduce additional bias. The current work presents a thorough investigation of these effects in primate brain.

**Methods:** Gradient-recalled echo (GRE) data of an entire fixed chimpanzee brain were acquired at 7 T (350 μm resolution, 10 orientations) including ideal COSMOS sampling and realistic rotations *in vivo*. Comparisons of the results included ideal COSMOS, *in-vivo* feasible acquisitions with 3–8 orientations and single-orientation iLSQR QSM.

**Results:** *In-vivo* feasible and optimal COSMOS yielded high-quality susceptibility maps with increased SNR resulting from averaging multiple acquisitions. COSMOS reconstructions from non-ideal rotations about a single axis required additional L2-regularization to mitigate residual streaking artifacts.

**Conclusion:** In view of unconsidered anisotropy effects, added complexity of the reconstruction, and the general challenge of multi-orientation acquisitions, advantages of sub-optimal COSMOS schemes over regularized single-orientation QSM appear limited in *in-vivo* settings.

## Keywords

Anisotropic magnetic susceptibility; COSMOS; Gradient-recalled echo; Quantitative susceptibility mapping.

# 1 | Introduction

Local alterations of iron and myelin in the brain are often linked to pathologies, making their quantitation an important goal for diagnostic and neuroscientific imaging. Given the distinct magnetic properties of iron-containing compounds (typically paramagnetic) and myelin (diamagnetic), alterations in their local content cause spatial variations of the bulk magnetic susceptibility $\Delta\chi$.[1] Quantification methods employing magnetic resonance imaging (MRI) to obtain local $\Delta\chi$ estimates have, therefore, been proposed, in particular, quantitative susceptibility mapping (QSM).[2-4] In QSM, susceptibility is estimated indirectly from frequency shifts measured by the gradient echo (GRE) signal phase $\varphi$.[5,6] The processing pipeline includes the estimation of the full phase evolution using phase-unwrapping techniques, the removal of background-field contributions from the local magnetic field variation $\delta B_0$ and, finally, a field-to-source inversion method to extract $\Delta\chi$ from $\delta B_0$ under the strict assumption that $\Delta\chi$ within a voxel is a scalar, isotropic quantity. Two basic issues arise for QSM: *(i)* The inverse problem from local tissue phase to susceptibility is intrinsically ill-posed because the dipole response kernel has near-zero values in the vicinity of two conical surfaces in the Fourier domain. Therefore, a direct inversion is ill conditioned at such positions, and hence, additional assumptions are required for avoiding artifacts in the final estimation. *(ii)* Magnetic susceptibility—especially in white matter (WM)—exhibits anisotropic characteristics as a result of the specific liquid-crystalline arrangement of elongated lipid molecules within the myelin sheaths enveloping axons. This inherent anisotropy is typically ignored in QSM pipelines.[6,7]

Most QSM implementations employ post-acquisition regularization solutions to the first problem, for example iLSQR-QSM.[8] Calculation Of Susceptibility through Multiple Orientation Sampling (COSMOS)[2] has been introduced as an acquisition-based solution. It requires sampling at three or more orientations of the object inside the magnetic field. An analytical investigation of the optimal scheme suggests that rotations by 0°, 60° and 120° about the magnet's (physical) *x*-axis eliminate the 'χάος'[1] of conic surfaces with zeroes and permit the estimation of $\Delta\chi$ without further regularization. With this choice of orientations, COSMOS fully addresses the ill-posed inverse problem yielding susceptibility maps that should be free of streaking artifacts. Considering the second problem, by sampling optimally along a circle, susceptibility anisotropy is partially averaged out (at least along the direction of sample rotation), leading to a more robust $\Delta\chi$ estimate. Results obtained with COSMOS have, therefore, been used previously as gold standard susceptibility reference in evaluations of QSM processing pipelines.[9] However, the application of COSMOS in human brain imaging *in vivo* is challenging because achievable head rotation angles are limited to a small range, for example ±25° or less,[10–14] which deviates substantially from the ideal rotation scheme. To address the limitation of a restricted range of accessible angles, head rotations about two axes have also been advocated as alternative to the optimal COSMOS scheme.[15]

---

[1] 'Chaos', in Greek mythology the void state preceding the creation of 'cosmos' (the universe).

In this work, we used high-resolution 7T GRE acquisitions in *post-mortem* chimpanzee brain as a versatile and robust setup for a further evaluation of COSMOS.[16] In particular, we compared the optimal scheme to results obtained with a more realistic range of accessible angles *in vivo* in the human brain and with single-orientation iLSQR-QSM. Acquisitions using more than three orientations were also included in these comparisons to evaluate the potential impact from more expanded orientation information (or averaging) in $\Delta\chi$ estimations. Finally, residual effects from susceptibility anisotropy on the obtained maps were evaluated.

## 2 | Methods

### *2.1 Brain specimen*

The specimen used for all acquisitions was obtained *post-mortem* from a female chimpanzee (*Pan troglodytes verus*) from Kolmården Wildlife Park, Sweden. The chimpanzee was medically euthanized due to a cervical leiomyoma and secondary hydronephrosis. All procedures followed the ethical guidelines of primatology research at the Max Planck Institute for Evolutionary Anthropology, Leipzig, and were approved by the Ethics Committee of the Max Planck Society. Exhaustive details of the steps involved in the brain extraction, fixation and storage have been published by Gräßle *et al*.[17] Briefly, a specialized veterinarian performed the brain extraction within a time interval of 18 h after death. The tissue was preserved in 4% paraformaldehyde (PFA) in phosphate-buffered saline (PBS). For the MRI acquisitions, the PFA was washed out in PBS at pH 7.4. The specimen was then immersed in Fomblin® (Solvay Solexis, Bollate, Italy) and positioned in an individualized, anatomically-shaped, 3D-printed container to avoid gravity-induced non-linear deformations during reorientation. Further details of this setup are given by Gkotsoulias *et al.*[18]

## 2.2 Multi-orientation phase measurements

Complex-valued 3D Fast Low-Angle SHot (FLASH)[19] data were obtained at 7 T on a MAGNETOM Terra (Siemens Healthineers, Erlangen Germany) with a single-channel transmit/32-channel receive head coil (Nova Medical, Wilmington, MA, USA). The acquisition parameters were optimized for high-resolution *post-mortem* acquisitions (field of view 160×160×145 mm³, matrix size 458×458×416, 0.35mm isotropic nominal resolution, transverse orientation) based on prior experience.[20,21] To account for $T_1$ and $T_2^*$ shortening in fixed tissue, a radiofrequency (RF) pulse flip angle of $\alpha = 28°$ was combined with a relatively short repetition time (*TR* = 26 ms) for efficient sampling (bandwidth 180 Hz/pixel), while the echo time of *TE* = 13 ms was long enough to achieve sufficient phase evolution. The phase encoding (PE) direction was from right to left (here, along the *z*-axis of the laboratory frame). GeneRalized Autocalibrating Partially Parallel Acquisitions (GRAPPA)[22] with acceleration factor *R* = 2 and a partial-Fourier scheme[23] with partial-Fourier factor, $f_p$ = 7/8 were employed in PE direction to accelerate the measurements (acquisition time, *TA* ≈ 34 min per orientation). The analytical scanning parameters are noted in Supplementary Material - Table ST2. In total, ten measurements were obtained, starting with the LPI reference (coordinate system with axis—in terms of the subject—from left to right, posterior to anterior, and inferior to superior) and followed by rotating the specimen by angles as given in Table 1 (also shown in Supplementary material - Figure SF1). After every two scans, the acquisition was paused for approximately 1 h to mitigate effects from potential temperature drifts due to the energy deposited in the specimen by repeated application of RF pulses.

## 2.3 Image processing

The uncombined image data from the individual receive channels were retrieved from the scanner using an in-house functor.[24,25] Multi-channel combination without phase singularities was accomplished using an adaptive-combine algorithm implemented in MATLAB (v. 2022.a; MathWorks, Natick, MA, USA), followed by Laplacian phase unwrapping[26] and V-SHARP background-field removal[27,28] for obtaining the tissue phase volumes of high quality. The phase data for each orientation were registered to the LPI reference employing transformations that were derived by registering the corresponding magnitude volumes using FMRIB's Linear Image Registration Tool (FLIRT; FSL 5.0.9) with six-parameter rigid transformations, a normalized Mutual Information (nMI) cost function and spline interpolation.[29] All registrations achieved excellent overlap, and the transformation matrices indicated angular transformations that were close to the experimentally adjusted ones. FSL was also used to derive a full-brain mask from the reference volume.

Quantitative susceptibility maps were reconstructed from the tissue phase of the reference orientation using iLSQR.[8] Using MATLAB, the remaining ten registered tissue phase volumes were used in different combinations for a total of six COSMOS reconstructions as indicated in Table 1. These combinations included the optimal COSMOS scheme ("COSMOS-opt" with rotations of $0°_x$, $60°_x$ and $120°_x$ about the *x*-axis), two combinations of three rotation angles mimicking the limited range of accessible rotation angles under conditions of *in-vivo* MRI of the human brain ("COSMOS-iv1" and "COSMOS-iv2" with rotations of $0°_x$, $10°_x$, $22°_x$ and of $0°_x$, $22°_x$, $-22°_x$ about the *x*-axis, respectively), and three further combinations to explore the potential advantage obtained with more than three orientations about either a single rotation axis ("COSMOS-iv3" with rotations of $0°_x$, $10°_x$, $22°_x$ and $-22°_x$ about the *x*-axis) or multiple rotations axes ("COSMOS-iv4" with rotations of $0°_x$, $10°_x$, $22°_x$, $-22°_x$ and of $22°_y$, $-22°_y$ about the *x*- and *y*-axis, respectively and "COSMOS-iv5" with rotations of $0°_x$, $10°_x$, $22°_x$, $-22°_x$, of $22°_y$, $-22°_y$ and of $45°_z$, $-45°_z$ about the *x*-, *y*- and *z*-axis, respectively). The scripts for COSMOS and L2-regularized COSMOS were modifications based on openly available MATLAB implementations (https://martinos.org/~berkin/software.html). The statistical analyses included the 3D structural similarity index measure (SSIM) and Pearson's correlation coefficient (*R*). The measures were computed from a large region of interest (ROI) consisting of gray matter (GM) and white matter (WM) as previously employed in a QSM reconstruction challenge.[9] Normalized difference maps were obtained by first normalizing the $\Delta\chi$ maps within the range [0,1] and then computing the absolute differences. The Gaussian noise variance was estimated in a 200×200×200 matrix extending from the center of the specimen.[30,31]

# 3 | Results

An example slice (magnitude and phase image) acquired with different specimen orientations and registered to the LPI reference is presented in Figure 1. Phase maps acquired with relatively small rotation angles (e.g., ±10°) are almost indistinguishable from the reference, whereas maps acquired with rotation angles exceeding 20° (e.g., 60° and 120° about the *x*-axis) show distinct deviations. Such differences are particularly prominent in WM regions, where relevant susceptibility anisotropy impacts the local field distribution. This suggests that COSMOS reconstructions from different sets of rotation angles will vary in regions of relevant susceptibility anisotropy. For example, scheme COSMOS-iv1 will be dominated by the measured phase distribution at the reference position (voxel-by-voxel averaging of similar local phase values), whereas the input images to the optimal scheme COSMOS-opt differ considerably in their phase variation (voxel-by-voxel averaging of local phase values distributed over a larger range). This is further evident from a comparison of the various COSMOS reconstructions with the simple QSM results obtained with iLSQR (Figure 2). All acquisitions with *in vivo*-feasible rotation angles about a single axis (e.g., *x*-axis) yielded COSMOS reconstructions with residual streaking artifacts. These artifacts were effectively mitigated by L2-regularization in all cases. The optimal COSMOS reconstruction and the *in vivo*-feasible schemes that included rotations about more than one axis appeared artifact-free without regularization.

As expected, the Gaussian noise variance was reduced for COSMOS reconstructions (range from $5 \times 10^{-4}$ to $8 \times 10^{-4}$) compared to iLSQR-QSM (approx. $10^{-3}$). Consistent with this observation, the use of only single orientations (i.e., no averaging) appeared to result in less smoothing. We note that the noise variance served as a (relative) metric in the current case as the calculation of the SNR is not straightforward for susceptibility maps.

Figures 3 presents normalized difference maps of results obtained with iLSQR-QSM and COSMOS-opt in comparison to corresponding difference maps obtained with iLSQR-QSM and COSMOS-iv2. As expected, prominent differences between iLSQR-QSM and COSMOS-opt are observed in WM. Differences between iLSQR-QSM and COSMOS-iv2 are also evident, however, they appear reduced. The enhanced similarity of the iLSQR-QSM and COSMOS-iv2 results (compared to COSMOS-opt) is particularly evident in WM regions of high anisotropy, such as corpus callosum or internal capsule,[7,18] whereas WM regions with more complex fiber dispersion or GM yielded higher agreement of the two COSMOS schemes. Note that the regularization that was included in the COSMOS-iv2 (but not in the COSMOS-opt) reconstruction also adds to these differences in a widespread fashion.

These observations are further corroborated by the SSIM and Pearson's correlation coefficient shown in Figure 4. Using COSMOS-opt as reference, the greatest SSIM and $R$ values were obtained with COSMOS-iv2 and schemes with more rotations. This is consistent with the expectations that larger rotation angles are beneficial for COSMOS reconstructions if the angles deviate from the optimal distribution. Acquisitions with further rotations (albeit without increasing the angle span) yielded similar SSIM and $R$ values. Compared to the various COSMOS schemes, iLSQR-QSM yielded the lowest SSIM and $R$ although its accuracy measures were only moderately lower than that obtained with *in vivo*-feasible COSMOS schemes (the exact values of the metrics results can be found in Supplementary Material - Table ST1).

## 4 | Discussion

Fixed brain specimens can be arbitrarily rotated, overcoming limitations of orientation-dependent *in vivo* experiments. The chimpanzee brain is smaller in size than a human brain, supporting less time-consuming high-resolution acquisitions while still providing a similarly complex, rich WM architecture. The accuracy of the rotation-angle adjustment was high (estimated error ≤2° based on the registration results), facilitated through a sophisticated setup that was designed to support reorientation experiments.[18] This is different from *in-vivo* scanning of the human brain, where multiple unstable factors impact the robustness of the rotations, such as the subject's head size and shape, fixation within the coil, or head motion. Close-fitting array coils are optimized for high SNR in the LPI position and do not well support reorientation of the head, which severely limits the range of accessible rotation angles for the majority of adult subjects.

The signal distribution inside the object depends on the specific distance from individual coil elements and might vary upon reorientations. Such inconsistencies lead to SNR fluctuations and inaccuracies in the registration and post-processing pipelines and may, hence, further degrade the quality of COSMOS results or other multi-orientation QSM methods. Indications of this bias is evident on the magnitude images after 60° and 120° rotations about the *x*-axis ([Figure 1](#)), where the signal intensity is reduced in the frontal lobe and increased in parieto-occipital brain compared to the LPI reference. This may require further preprocessing before registration, such as N4 bias correction,[32] which was not employed in our experiments.

Our results indicate limited performance of the COSMOS model when applied to data with a restricted range of rotation angles as typically encountered *in-vivo*. While the results from suboptimal angle distributions still outperformed single-orientation QSM in comparisons to the optimal COSMOS scheme, the similarity metrics SSIM and *R* for the full-brain data were only <0.9 and <0.85, respectively. Qualitatively, this is consistent with previous work indicating slightly better performance for multi-angle QSM (with a restricted range of angles) compared to both thresholded or regularized single-angle approaches.[10] We note that our data were recorded under idealized conditions, whereas subtle head motion *in vivo* might degrade multi-orientation acquisitions (requiring longer scan time) more than those with only a single orientation. Extending the rotations beyond one axis relaxed the need for regularization to eliminate streaking artifacts.

An inherent limitation is that COSMOS does not consider susceptibility anisotropy. This aspect has only recently received more attention,[33] whereas previous investigations were typically limited to deep GM structures.[10,14] Anisotropy leads to a variation of the local field distribution upon reorienting the sample in the magnetic field. Overall, the optimal COSMOS scheme (0°, –60°, –120°) appeared to achieve better averaging of anisotropic susceptibility, especially in parts of WM with strong orientational characteristics. In comparison, averaging of anisotropy is limited with the restricted range of angles *in-vivo*, resulting in enhanced orientation bias. Apart from such performance differences, all COSMOS reconstructions will be impacted by anisotropy-related variability, in particular in WM regions of higher order of the local fiber distributions ([Figure 3](#)). Remarkably, the previously suggested $\chi_{33}$ component of the susceptibility tensor[9] does not appear to be a better gold standard for scalar QSM estimations, presumably due to unaccounted microstructural effects contained in off-diagonal susceptibility tensor elements.[33] A better approach to the ground truth may be obtained by projecting the $\chi_{13}$ and $\chi_{23}$ anisotropic contributions into the apparent scalar susceptibility, which is, however, not trivial.

For COSMOS reconstructions from non-optimal angle distributions limited to a single rotation axis (e.g., the $x$-axis), additional L2-regularization was required to mitigate residual streaking artifacts. This is counter-intuitive because COSMOS was introduced as a method for solving the ill-posed inversion problem without regularization—as indeed achieved with COSMOS-opt or upon incorporating rotations about multiple axes. Overall, both *in-vivo* feasible and optimal COSMOS yielded high-quality susceptibility maps with increased SNR resulting from inherent averaging of multiple acquisitions. Nevertheless, in view of unconsidered anisotropy effects, added complexity of the reconstruction, and the general challenge of multi-orientation acquisitions, advantages of sub-optimal COSMOS schemes over regularized single-orientation QSM appear limited.

## Conflict of interest

The authors declare no competing interest.

## Data availability statement

The data that support the findings of this study are available from the corresponding author or the senior author upon reasonable request.

## Acknowledgements

This work was funded by a grant to HEM from the European Commission, ITN "INSPiRE-MED" (H2020-MSCA-ITN-2018, #813120). BB acknowledges funding by the National Institutes of Health, grant Nos. R01 EB032378, R01 EB028797, R03 EB031175, U01 EB025162, P41 EB030006, and U01 EB026996, and computing support by the NVIDIA Corporation. We

are grateful to Angela D. Friederici and Nikolaus Weiskopf from the Max Planck Institute for Human Cognitive and Brain Sciences and all Principal Investigators of the Evolution of Brain Connectivity (EBC) project for supporting this research through providing access to the brain specimen. We especially thank Ariane Düx and Fabian Leenderz from the Helmholtz Institute for One Health for their involvement in establishing the brain extraction network and the Kolmården Wildlife Park in Kolmården and National Veterinary Institute, Sweden, for executing the brain extraction and providing the tissue.## References

1. Möller HE, Bossoni L, Connor JR, Crichton RR, Does MD, Ward RJ, Zecca L, Zucca FA, Ronen I. Iron, myelin, and the brain: Neuroimaging meets neurobiology. *Trends Neurosci.* 2019; 42: 384–401. https://doi.org/10.1016/j.tins.2019.03.009.

2. Liu T, Spincemaille P, de Rochefort L, Kressler B, Wang Y. Calculation of susceptibility through multiple orientation sampling (COSMOS): A method for conditioning the inverse problem from measured magnetic field map to susceptibility source image in MRI. *Magn. Reson. Med.* 2009; 61: 196–204. https://doi.org/10.1002/mrm.21828.

3. Shmueli K, de Zwart JA, van Gelderen P, Li T-Q, Dodd SJ, Duyn JH. Magnetic susceptibility mapping of brain tissue in vivo using MRI phase data. *Magn. Reson. Med.* 2009; 62: 1510–1522. https://doi.org/10.1002/mrm.22135.

4. Schweser F, Deistung A, Lehr BW, Reichenbach JR. Quantitative imaging of intrinsic magnetic tissue properties using MRI signal phase: An approach to in vivo brain iron metabolism? *NeuroImage* 2011; 54: 2789–2807. https://doi.org/10.1016/j.neuroimage.2010.10.070.

5. Liu C, Wei H, Gong N-J, Cronin M, Dibb R, Decker K. Quantitative susceptibility mapping: Contrast mechanisms and clinical applications. *Tomography* 2015; 1: 3–17. https://doi.org/10.18383/j.tom.2015.00136.

6. Deistung A, Schweser F, Reichenbach JR. Overview of quantitative susceptibility mapping. *NMR Biomed.* 2017; 30: e3569. https://doi.org/10.1002/nbm.3569.

7. Liu C. Susceptibility tensor imaging. *Magn. Reson. Med.* 2010; 63: 1471–1477. https://doi.org/10.1002/mrm.22482.

8. Li W, Wang N, Yu F, Han H, Cao W, Romero R, Tantiwongkosi B, Duong TQ, Liu C. A method for estimating and removing streaking artifacts in quantitative susceptibility mapping. *NeuroImage* 2015; 108: 111–122. https://doi.org/10.1016/j.neuroimage.2014.12.043.

9. Langkammer C, Schweser F, Shmueli K, Kames C, Li X, Guo L, Milovic C, Kim J, Wie H, Bredies K, Buch S, Guo Y, Liu Z, Meineke J, Rauscher A, Marques JP, Bilgic B. Quantitative susceptibility mapping: Report from the 2016 reconstruction challenge. *Magn. Reson. Med.* 2018; 79: 1661–1673. https://doi.org/10.1002/mrm.26830.

10. Wharton S, Bowtell R. Whole-brain susceptibility mapping at high field: A comparison of multiple- and single-orientation methods. *NeuroImage* 2010; 53: 515–525. https://doi.org/10.1016/j.neuroimage.2010.06.070.

11. Schweser F, Sommer K, Deistung A, Reichenbach JR. Quantitative susceptibility mapping for investigating subtle susceptibility variations in the human brain. *NeuroImage* 2012; 62: 2083–2100. https://doi.org/10.1016/j.neuroimage.2012.05.067.

12. Yoon J, Gong E, Chatnuntawech I, Bilgic B, Lee J, Jung W, Ko J, Jung H, Setsompop K, Zaharchuk G, Kim EY, Pauly J, Lee J. Quantitative susceptibility mapping using deep neural network: QSMnet. *NeuroImage* 2018; 179: 199–206. https://doi.org/10.1016/j.neuroimage.2018.06.030.

13. Liu T, Liu J, de Rochefort L, Spincemaille P, Khalidov I, Ledoux JR, Wang Y. Morphology enabled dipole inversion (MEDI) from a single-angle acquisition: Comparison with COSMOS in human brain imaging. *Magn. Reson. Med.* 2011; 66: 777–783. https://doi.org/10.1002/mrm.22816.

14. Bilgic B, Xie L, Dibb R, Langkammer C, Mutluay A, Ye H, Polimeni JR, Augustinack J, Liu C, Wald LL, Setsompop K. Rapid multi-orientation quantitative susceptibility mapping. *NeuroImage* 2016; 125: 1131–1141. https://doi.org/10.1016/j.neuroimage.2015.08.015.

15. Wharton S, Schäfer A, Bowtell R. Susceptibility mapping in the human brain using threshold-based *k*-space division. *Magn. Reson. Med.* 2010; 63: 1292–1304. https://doi.org/10.1002/mrm.22334.

16. Gkotsoulias DG, Müller R, Schlumm T, Alsleben N, Jäger C, Jaffe J, Pampel A, Crockford C, Wittig R, Möller HE. COSMOS-based susceptibility estimations: Accuracy assessment and comparisons of QSM and multiple-angle acquisitions. *In:* Proceedings of the 31st Annual Meeting of ISMRM, London, UK, 2022, p. 2370.

17. Gräßle T, Crockford C, Eichner C, Girard-Buttoz C, Jäger C, Kirilina E, Lipp I, Düx A, Edwards L, Jauch A, Kopp KS, Paquette M, Pine K, EBC Consortium, Haun DBM, McElreath R, Anwander A, Gunz P, Morawski M, Friederici AD, Weiskopf N, Leendertz FH, Wittig RM. Sourcing high tissue quality brains from deceased wild primates with known socio-ecology. *Methods Ecol. Evol.* 2023; 14: 1906–1924. https://doi.org/10.1111/2041-210X.14039.

18. Gkotsoulias DG, Müller R, Jäger C, Schlumm T, Mildner T, Eichner C, Pampel, A, Jaffe J, Gräßle T, Alsleben N, Chen J, Crockford C, Wittig R, Liu C, Möller HE. High angular resolution susceptibility imaging and estimation of fiber orientation distribution functions in primate brain. *NeuroImage* 2023; 276: 120202. https://doi.org/10.1016/j.neuroimage.2023.120202.

19. Frahm J, Haase A, Matthaei D. Rapid three-dimensional MR imaging using the FLASH technique. *J. Comput. Assist. Tomogr.* 1986; 10: 363–368. https://doi.org/10.1097/00004728-198603000-00046.

20. Alkemade A, Pine K, Kirilina E, Keuken MC, Mulder MJ, Balesar R, Groot JM, Bleys RLAW, Trampel R, Weiskopf N, Herrler A, Möller HE, Bazin P-L, Forstmann BU. 7 Tesla MRI followed by histological 3D reconstructions in whole-brain specimens. *Front. Neuoanat.* 2020; 14: 536838. https://doi.org/10.3389/fnana.2020.536838.

21. Alkemade A, Bazin P-L, Balesar R, Pine K, Kirilina E, Möller HE, Trampel R, Kros JM, Keuken MC, Bleys RLAW, Swaab DF, Herrler A, Weiskopf N, Forstmann BU. A unified 3D map of microscopic architecture and MRI of the human brain. *Sci. Adv.* 2022; 8: eabj7892. https://doi.org/10.1126/sciadv.abj7892.

22. Griswold MA, Jakob PM, Heidemann RM, Nittka M, Jellus V, Wang J, Kiefer B, Haase A. Generalized autocalibrating partially parallel acquisitions (GRAPPA). *Magn. Reson. Med.* 2002; 47: 1202–1210. https://doi.org/10.1002/mrm.10171.

23. Feinberg DA, Hale JD, Watts JC, Kaufman L, Mark A. Halving MR imaging time by conjugation: Demonstration at 3.5 kG. *Radiology* 1986; 161: 527–531. https://doi.org/10.1148/radiology.161.2.3763926.

24. Hellrung L, Dietrich A, Maurice Hollmann M, Pleger B, Kalberlah C, Roggenhofer E, Villringer A, Horstmann A. Intermittent compared to continuous real-time fMRI neurofeedback boosts control over amygdala activation. *NeuroImage* 2018; 166: 198–208. https://doi.org/10.1016/j.neuroimage.2017.10.031.

25. Kanaan AS, Yu D, Metere R, Schäfer A, Schlumm T, Bilgic B, Anwander A, Mathews CA, Scharf JM, Müller-Vahl K, Möller HE. Convergent imaging-transcriptomic evidence for disturbed iron homeostasis in Gilles de la Tourette syndrome. *Neurobiol. Dis.* 2023. https://doi.org/10.1016/j.nbd.2023.106252.

26. Schofield MA, Zhu Y. Fast phase unwrapping algorithm for interferometric applications. *Opt. Lett.* 2003; 28: 1194–1196. https://doi.org/10.1364/OL.28.001194.

27. Li W, Wu B, Liu C. Quantitative susceptibility mapping of human brain reflects spatial variation in tissue composition. *NeuroImage* 2011; 55: 1645–1656. https://doi.org/10.1016/j.neuroimage.2010.11.088.

28. Özbay PS, Deistung A, Feng X, Nanz D, Reichenbach JR, Schweser F. A comprehensive numerical analysis of background phase correction with V-SHARP. *NMR Biomed.* 2017; 30: e3550. https://doi.org/10.1002/nbm.3550.

29. Jenkinson M, Beckmann CF, Behrens TEJ, Woolrich MW, Smith SM. FSL. *NeuroImage* 2012; 62: 782–790. https://doi.org/10.1016/j.neuroimage.2011.09.015.

30. Immerkær J. Fast noise variance estimation. *Comput. Vis. Image Underst.* 1996; 64: 300–302. https://doi.org/10.1006/cviu.1996.0060.

31. Liu W, Liu T, Rong M, Wang R, Zhang H. A fast noise variance estimation algorithm. *In:* Proceedings of the 2011 Asia Pacific Conference on Postgraduate Research in Microelectronics & Electronics, Macao, China, 2011, pp. 61–64. https://doi.org/10.1109/PrimeAsia.2011.6075071.

32. Tustison NJ, Avants BB, Cook PA, Zheng Y, Egan A, Yushkevich PA, Gee JC. N4ITK: Improved N3 bias correction. *IEEE Trans Med Imaging* 2010; 29: 1310–1320. https://doi.org/10.1109/TMI.2010.2046908.

33. Milovic C, Özbay PS, Irarrazaval P, Tejos C, Acosta-Cabronero J, Schwesser F, Marques JP, Bilgic B, Langkammer C. The 2016 QSM Challenge: Lessons learned and considerations for a future challenge design. *Magn. Reson. Med.* 2020; 84: 1624–1637. https://doi.org/10.1002/mrm.28185.

# Figure captions

**Figure 1.** Arbitrarily selected axial slice of the registered magnitude images (top row) acquired with different specimen orientations. The corresponding phase maps (2$^{nd}$ row) show obvious inconsistencies, presumably resulting from the anisotropic properties of magnetic susceptibility. In the bottom row, the iLSQR QSM calculated for each orientation is presented. Note the significant changes in WM regions -especially in high rotations- presumably resulting from the anisotropic properties of magnetic susceptibility.

**Figure 2.** Susceptibility maps in a central coronal slice obtained with iLSQR-QSM as well as different COSMOS schemes including COSMOS-opt and those with rotations that are feasible under *in-vivo* conditions. The results obtained with COSMOS-iv2 and COSMOS-iv3 appear to have the greatest similarity to those from COSMOS-opt at visual inspection despite remaining differences in WM regions. Note that the results from all *in vivo*-feasible COSMOS acquisitions are also presented after L2-regularization (bottom row) to minimize potential residual streaking artifacts. The reconstructions from COSMOS-opt and from *vivo*-feasible acquisitions that included rotations about more than one axis appeared artifact-free without regularization.

**Figure 3.** Normalized difference maps of results from **(A)** iLSQR-QSM and COSMOS-opt and from **(B)** iLSQR-QSM and COSMOS-iv2 (indicated as "*in-vivo* COSMOS"). Zoomed ROIs are shown for better appreciation of the differences in specific parts of the brain. Pronounced deviations between iLSQR-QSM and COSMOS-opt in WM (e.g., corpus callosum and internal capsule) are less evident between iLSQR-QSM and COSMOS-iv2 in the same regions. **(C)** Fractional anisotropy (FA) and FA-weighted primary eigenvector (same ROIs as above) from registered diffusion tensor imaging acquired on the same specimen, in a separate scanning session at a different time (Red color denotes Left-Right direction of fibers; Blue color denotes Superior-Inferior direction; Green color denotes Anterior-Posterior direction, always in the object's coordinate system). Increased diffusion anisotropy regions is evident in regions of enhanced susceptibility differences supporting the assumption that susceptibility anisotropy is related to myelination in these WM regions.

**Figure 4**. Structural similarity index measure **(A)** and Pearson's correlation coefficient **(B)** as metrics for similarity and correlation of results obtained with iLSQR-QSM and *in vivo*-feasible COSMOS schemes in comparison to COSMOS-opt. Results are shown for masks [shown in blue color in **(C)**] of total GM (gray squares), total WM (open squares) and the combination of both (blue squares). COSMOS-iv2, yielded the smallest deviations than COSMOS-iv1 (i.e., greater SSIM and *R* values). Adding acquisitions with further rotations did not significantly improve the result. Compared to all COSMOS schemes, iLSQR-QSM yielded the greatest deviation (smallest SSIM and *R* values) from the COSMOS-opt result**.**

# Tables

**Table 1.** Overview of the sample rotations and the combinations of orientations used for the COSMOS reconstructions. The angles are indicated in degrees, and the *x*-, *y*-, and *z*-axis notation refers to the scanner coordinate system. The reference orientation (#1) was acquired with the specimen in the typical LPI *in-vivo* position. All other reoriented volumes were registered to this orientation.

|  | **1 (LPI)** | **2** | **3** | **4** | **5** | **6** | **7** | **8** | **9** | **10** |
|---|---|---|---|---|---|---|---|---|---|---|
| ***x*-axis rotation** | 0° | 60° | 120° | 10° | 22° | –22° | 0° | 0° | 0° | 0° |
| ***y*-axis rotation** | 0° | 0° | 0° | 0° | 0° | 0° | 22° | –22° | 0° | 0° |
| ***z*-axis rotation** | 0° | 0° | 0° | 0° | 0° | 0° | 0° | 0° | 45° | –45° |
| **COSMOS-opt** | ✓ | ✓ | ✓ |  |  |  |  |  |  |  |
| **COSMOS-iv1** | ✓ |  |  | ✓ | ✓ |  |  |  |  |  |
| **COSMOS-iv2** | ✓ |  |  |  | ✓ | ✓ |  |  |  |  |
| **COSMOS-iv3** | ✓ |  |  | ✓ | ✓ | ✓ |  |  |  |  |
| **COSMOS-iv4** | ✓ |  |  | ✓ | ✓ | ✓ | ✓ | ✓ |  |  |
| **COSMOS-iv5** | ✓ |  |  | ✓ | ✓ | ✓ | ✓ | ✓ | ✓ | ✓ |

# Figures

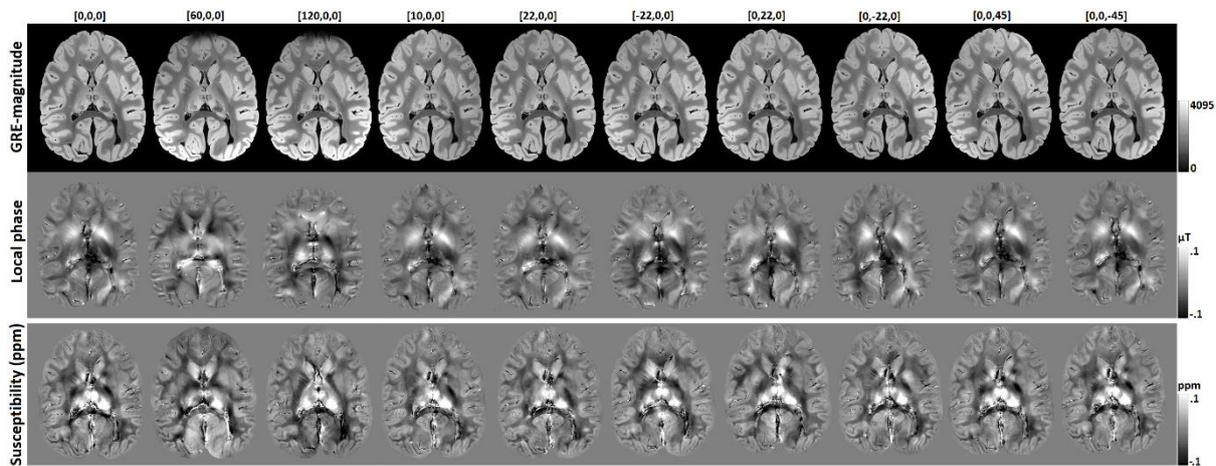

**Figure 1.** Arbitrarily selected axial slice of the registered magnitude images (top row) acquired with different specimen orientations. The corresponding phase maps (2nd row) show obvious inconsistencies, presumably resulting from the anisotropic properties of magnetic susceptibility. In the bottom row, the iLSQR QSM calculated for each orientation is presented. Note the significant changes in WM regions -especially in high rotations- presumably resulting from the anisotropic properties of magnetic susceptibility.

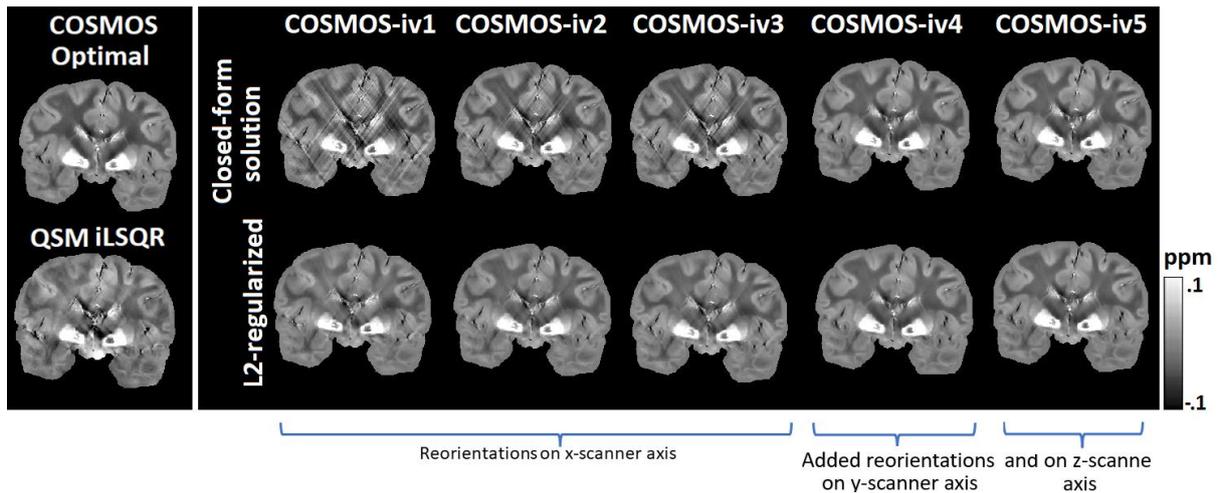

**Figure 2.** Susceptibility maps in a central coronal slice obtained with iLSQR-QSM as well as different COSMOS schemes including COSMOS-opt and those with rotations that are feasible under *in-vivo* conditions. The results obtained with COSMOS-iv2 and COSMOS-iv3 appear to have the greatest similarity to those from COSMOS-opt at visual inspection despite remaining differences in WM regions. Note that the results from all *in vivo*-feasible COSMOS acquisitions are also presented after L2-regularization (bottom row) to minimize potential residual streaking artifacts. The reconstructions from COSMOS-opt and from *vivo*-feasible acquisitions that included rotations about more than one axis appeared artifact-free without regularization.

**Figure 3.** Normalized difference maps of results from **(A)** iLSQR-QSM and COSMOS-opt and from **(B)** iLSQR-QSM and COSMOS-iv2 (indicated as "*in-vivo* COSMOS"). Zoomed ROIs are shown for better appreciation of the differences in specific parts of the brain. Pronounced deviations between iLSQR-QSM and COSMOS-opt in WM (e.g., corpus callosum and internal capsule) are less evident between iLSQR-QSM and COSMOS-iv2 in the same regions. **(C)** Fractional anisotropy (FA) and FA-weighted primary eigenvector (same ROIs as above) from registered diffusion tensor imaging acquired on the same specimen, in a separate scanning session at a different time (Red color denotes Left-Right direction of fibers; Blue color denotes Superior-Inferior direction; Green color denotes Anterior-Posterior direction, always in the object's coordinate system). Increased diffusion anisotropy regions is evident in regions of enhanced susceptibility differences supporting the assumption that susceptibility anisotropy is related to myelination in these WM regions.

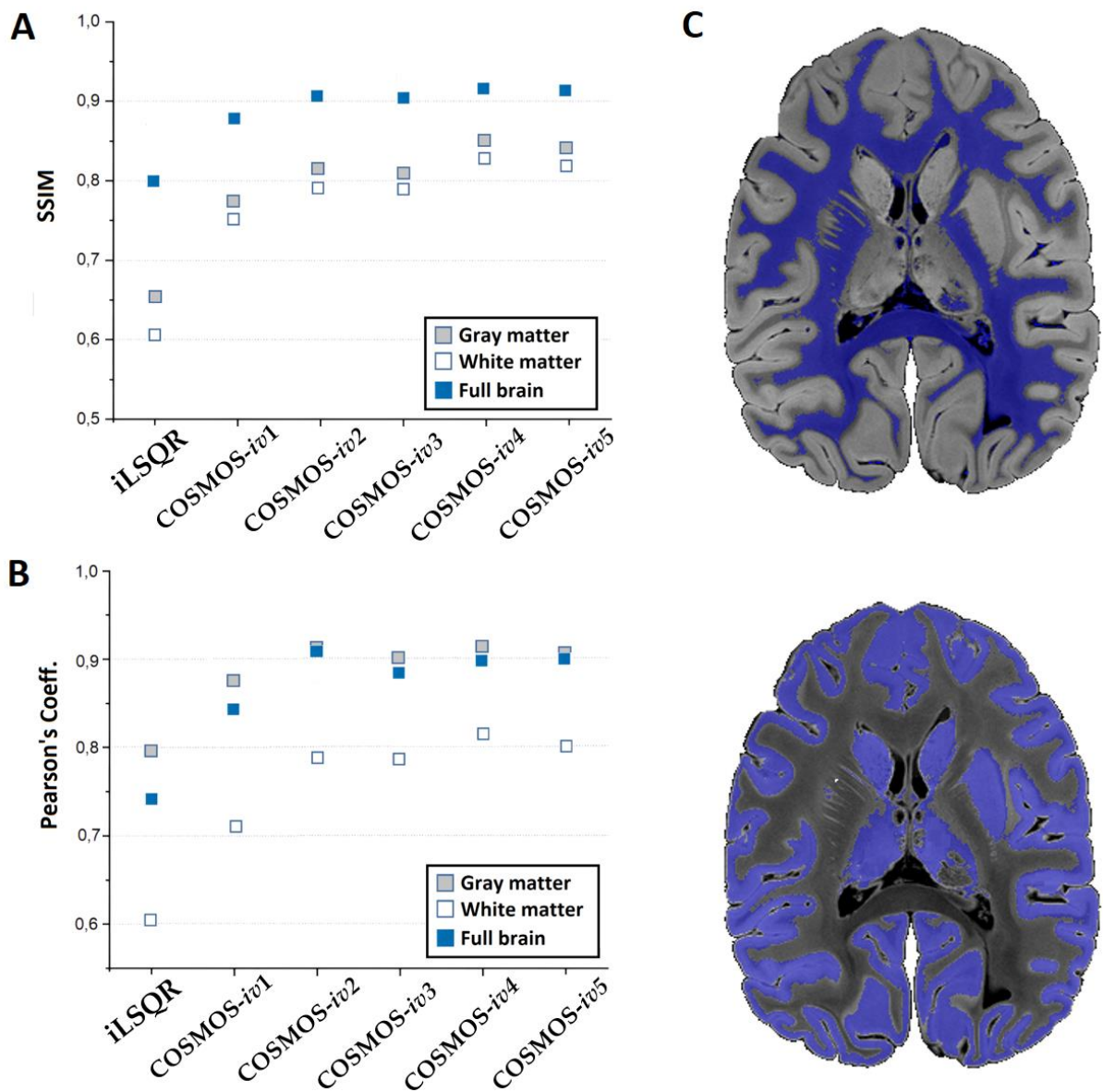

**Figure 4**. Structural similarity index measure **(A)** and Pearson's correlation coefficient **(B)** as metrics for similarity and correlation of results obtained with iLSQR-QSM and *in vivo*-feasible COSMOS schemes in comparison to COSMOS-opt. Results are shown for masks [shown in blue color in **(C)**] of total GM (gray squares), total WM (open squares) and the combination of both (blue squares). COSMOS-iv2, yielded the smallest deviations than COSMOS-iv1 (i.e., greater SSIM and *R* values). Adding acquisitions with further rotations did not significantly improve the result. Compared to all COSMOS schemes, iLSQR-QSM yielded the greatest deviation (smallest SSIM and *R* values) from the COSMOS-opt result.